\documentclass{aa}  
\usepackage{graphicx}
\usepackage{txfonts}
\begin{document}

\title{Brown dwarf formation by binary disruption}

\author{Simon P. Goodwin \inst{1} \and Ant Whitworth \inst{2}}

\offprints{Simon Goodwin}

\institute{Department of Physics \& Astronomy, University of
  Sheffield, Hicks Building, Hounsfield Road, Sheffield, S3 7RH, UK.\\
\email{S.Goodwin@sheffield.ac.uk} 
\and School of Physics \& Astronomy, Cardiff University, The Queens 
Buildings, 5 The Parade, Cardiff, CF24 3AA, Wales, UK.\\
\email{A.Whitworth@astro.cf.ac.uk}}

\date{Received November 15, 2006; accepted }

\abstract
{The principal mechanism by which brown dwarfs form, and its relation
to the formation of higher-mass (i.e. hydrogen-burning) stars, is
poorly understood.}
{We advocate a new model for the formation of brown dwarfs.}
{In this model, brown dwarfs are initially binary companions, formed
by gravitational fragmentation of the outer parts ($R\ga 100\,{\rm
AU}$) of protostellar discs around low-mass hydrogen-burning
stars. Most of these binaries are then gently disrupted by passing
stars to create a largely single population of brown dwarfs and
low-mass hydrogen-burning stars.}
{This idea is consistent with the excess of binaries found in
low-density pre-main sequence populations, like that in Taurus, where
they should survive longer than in denser clusters. }
{If brown dwarfs form in this way, as companions to more massive
stars, the difficulty of forming very low-mass prestellar cores is
avoided. Since the disrupted binaries will tend to be those involving
low-mass components and wide orbits, and since disruption will be due
to the gentle tides of passing stars (rather than violent $N$-body
interactions in small-$N$ sub-clusters), the liberated brown dwarfs
will have velocity dispersions and spatial distributions very similar
to higher-mass stars, and they will be able to retain discs, and
thereby to sustain accretion and outflows. Thus the problems
associated with the ejection and turbulence mechanisms can be
avoided.  This model implies that most, possibly
all, stars and brown dwarfs form in binary or multiple systems.}


\maketitle

\section{Introduction}\label{SEC:INTRO}

Brown dwarfs were first observed unambiguously over ten years ago
(Rebolo et al. 1995; Nakajima et al. 1995; Oppenheimer et
al. 1995). In the subsequent decade it has become clear that brown
dwarfs are very numerous, possibly comparable in number to
hydrogen-burning stars (Luhman et al. 2007). However, it is still not
clear how brown dwarfs form, nor how brown dwarfs relate to planets
and hydrogen-burning stars.

The two most frequently invoked theories for brown dwarf formation are
`ejection', where a stellar embryo is ejected from a small-N cluster,
before it can grow above $0.075\,{\rm M}_{_\odot}$ by accretion from
its natal core (e.g. Reipurth \& Clarke 2001; Bate et al. 2002;
Goodwin et al. 2004a,b; Delgado Donate et al. 2004) and `turbulence',
where very low-mass prestellar cores are formed by convergent flows
and collapse to form brown dwarfs in isolation (e.g. Padoan \&
Nordlund 2002, 2004). A number of other brown dwarf formation
mechanisms have been proposed, for example disc fragmentation (Boffin
et al. 1998; Watkins et al. 1998a,b; Whitworth \& Stamatellos 2006),
and photo-erosion of high-mass cores overrun by H{\sc ii} regions
(Whitworth \& Zinnecker 2004).  However, `ejection' and `turbulence'
are the two most popular theories (see Whitworth \& Goodwin 2005, and
Whitworth et al. 2007, for reviews of brown dwarf formation theory).

In this paper we review the observational properties of brown dwarfs
and low-mass hydrogen-burning stars, and rehearse some of the problems
associated with the two main formation theories. We then present a new
theory of brown dwarf formation which avoids these problems.

\section{The properties of brown dwarfs and low-mass hydrogen-burning 
stars}\label{SEC:PROPS}

There appears to be no good reason why brown dwarfs  
should form in a
different way from low-mass hydrogen-burning stars, and this
contention is supported by the observation that the statistical
properties of brown dwarfs form a continuum with those of low-mass
hydrogen-burning stars, across the divide at $0.075\,{\rm
M}_{_\odot}$. For example, the stellar Initial Mass Function is
continuous across the divide. It peaks at $M_{_\star}\sim0.3\,{\rm
M}_{_\odot}$ (Kroupa 2002; Chabrier 2003), and at least 20\% of
greater than planetary mass objects   
are brown dwarfs (Luhman et al. 2007). Similarly, the binary
statistics of brown dwarfs form a continuum with those of low-mass
hydrogen-burning stars: as primary mass is decreased across the
divide, the binary fraction decreases, the mean semi-major axis and
the logarithmic range of semi-major axes decrease, and the mean mass
ratio increases towards unity (Burgasser et al. 2006). In clusters,
the radial velocity dispersions and spatial distributions of brown dwarfs 
and low-mass hydrogen-burning stars are indistinguishable
(Brice\~no et al. 2002). Finally, brown dwarfs are observed to have
infrared excesses, indicative of discs (Muench et al. 2001; Natta \&
Testi 2001; Jayawardhana et al. 2003; Mohanty et al. 2004; Scholz et
al. 2006) and emission lines indicative of both on-going accretion
(e.g. Scholz \& Eisl\"offel 2004) and outflows (e.g. Fern\'andez \&
Comer\'on 2001; Natta et al. 2004; Whelan et al. 2005), just like
hydrogen-burning stars; the ratio of disc mass to stellar mass appears
to be approximately constant, and the accretion rate appears to scale
with stellar mass squared across the divide (Muzerolle et al. 2003,
2005).

\section{`Ejection' versus `turbulence'}\label{SEC:VERS}

Whitworth et al. (2007) have stressed that all proposed brown dwarf
  formation mechanisms probably produce at least some brown dwarfs.
  They emphasise that it is important to examine what fraction of
  brown dwarfs are formed by each mechanism.  In this section we argue
  that the two most popular mechanisms -- `ejection' and `turbulence'
  -- both face significant problems as the dominant mode of brown dwarf 
  formation.  These problems then lead us in the next section to
  propose a somewhat different mechanism which we believe may make a 
significant contribution to  brown dwarf
  formation. 

\subsection{Ejection}

Reipurth \& Clarke (2001) have proposed that brown dwarfs are formed
when a dense core collapses and fragments to form a small-$N$
sub-cluster, and then one (or more) of the resulting protostellar
embryos is ejected from the natal core, before it can accrete enough
mass to reach the hydrogen-burning limit. Once ejected, an embryo is
detached from the reservoir of material it might otherwise have
accreted, and its mass is essentially constant. In this mechanism, the
cores which spawn brown dwarfs must have sufficient mass to also
spawn two or three other stars. Therefore it is not necessary to
invoke the formation of very low-mass prestellar cores. Low-mass
hydrogen burning stars will be ejected from cores in the same way, but
since the probability of ejection decreases with increasing mass,
brown dwarfs will be ejected from cores preferentially and/or
sooner. It seems inescapable that this mechanism operates in
nature. However, there are a number of observations which suggest that
it may not be the dominant formation mechanism for brown dwarfs.

First, if brown dwarfs and low-mass hydrogen-burning stars are formed
by the ejection mechanism, they should have a larger velocity
dispersion -- and hence also in time a wider spatial extent -- than
more massive stars born in the same cluster. The effect is not
necessarily as dramatic as originally predicted by Reipurth \& Clarke
(2001), because typical ejection velocities are at most a few ${\rm
km}\,{\rm s}^{-1}$, and therefore comparable with the velocity
dispersion between neighbouring cores and the velocity dispersion for
more massive stars (Bate et al. 2003; Goodwin et
al. 2004a,b). However, there should be a high-velocity tail to the
distribution of ejection velocities, yielding an extended halo of
brown dwarfs, and this is not seen in Taurus (Brice{\~n}o et al. 2002;
Luhman 2006) or Chamaeleon (Joergens 2006).

Second, it appears that a significant fraction of brown dwarfs are
attended by  massive, extended discs, which sustain long-lived
accretion and outflows, just like more massive stars (Muench et
al. 2001; Natta \& Testi 2001; Muzerolle et al. 2003; Scholz \&
Eisloeffel 2004; Natta et al. 2004; Scholz et al. 2006). This is hard
to reconcile with the ejection mechanism. In simulations of core
collapse and fragmentation, ejected brown dwarfs do occasionally
retain significant discs (see Whitworth et al. 2007), but too
infrequently to explain the observations.

Third, the ejection of brown dwarfs and low-mass hydrogen-burning
stars quickly hardens the system of stars remaining near the centre of
the natal core. Simulations suggest that the collapse and
fragmentation of a core usually leads to the formation of a small-$N$
sub-cluster; and then ejection of the lower-mass members of the
sub-cluster typically reduces it rather quickly to a binary system,
often with quite a small separation, $\la 10\,{\rm AU}$ (e.g. Goodwin
et al. 2004a,b; Goodwin \& Kroupa 2005; Umbreit et al. 2005). As a
consequence of this early hardening, the binary system then normally
evolves towards equal-mass components. This is because the material
accreting later onto the binary system tends to have increasing
angular momentum, and therefore it can be accommodated more easily by
a low-mass secondary, which pushes the mass ratio, $q$, up towards
unity (Whitworth et al. 1995).  Whilst close companions are more likely
  to be of more equal mass (e.g. Duquennoy \& Mayor 1991; Mazeh et
  al. 1992), the observed trend is not as extreme as predicted by 
the standard  ejection scenario (see Goodwin et al. 2004b; Goodwin 
  et al. 2007).

\subsection{Turbulence}

Padoan \& Nordlund (2002, 2004) have proposed that brown dwarfs are
formed from the collapse of very low-mass prestellar cores, i.e. a
scaled-down version of the formation of more massive hydrogen-burning
stars (Andr\'e et al. 2000; Ward-Thompson et al. 2007). This proposal
is supported by the fact that the mass function for prestellar cores
appears to echo the shape of the stellar initial mass function, at
least at high and intermediate masses (Motte et al. 1998, 2001; Testi
\& Sargent 1998; Johnstone 2000; Nutter \& Ward-Thompson 2007). From
this it is presumed that the collapse and fragmentation of a
prestellar core having mass $M_{_{\rm CORE}}$ proceeds in a
statistically self-similar manner, spawning the same mix of scaled
stellar masses ($M_{_\star}/M_{_{\rm CORE}}$) with the same efficiency
($\sum\{M_{_\star}\}/M_{_{\rm CORE}}$).

If most prestellar cores spawn just one star (e.g. Lada 2006), the
shift of the peak -- from $1\,{\rm M}_{_\odot}$ for the prestellar
core mass function (Nutter \& Ward-Thompson 2007) to $0.3\,{\rm
M}_{_\odot}$ for the stellar Initial Mass Function (Chabrier 2003) --
implies a mean efficiency of $\sim\!30\%$ for converting prestellar
cores into stars. To form an isolated brown dwarf with mass
$M_{_\star}\sim 0.03\,{\rm M}_{_\odot}$ then requires a prestellar
core with mass $M_{_{\rm CORE}}\sim 0.1{\rm M}_{_\odot}$, and hence
diameter $D_{_{\rm CORE}} \sim 300\,{\rm AU}\,(T/10\,{\rm K})^{-1}$
and column-density $N_{_{\rm CORE}}\sim 2\times10^{24}\,{\rm
cm}^{-2}$. To form such a prestellar core in isolation requires an
implausibly well focussed and dense convergent flow, {\it viz.} gas
with density $n\sim 10^4\,{\rm cm}^{-3}$ flowing inwards (a)
inertially at speed $v\sim 1\,{\rm km}\,{\rm s}^{-1}$ and (b)
approximately isotropically over a sphere of radius $r\sim 0.01\,{\rm
pc}$, for a time $t\sim 10^5\,{\rm years}$ (Whitworth et al. 2007). In
an interstellar medium where the principal sources of turbulent energy
are shear and expansion, it seems very unlikely that these conditions
are fulfilled frequently enough to be a major source of brown dwarfs. More 
fundamentally, the pure turbulence theory predicts
explicitly that for every prestellar core with mass $M_{_{\rm
CORE}}\sim0.1{\rm M}_{_\odot}$ there are $\sim 20,000$ more diffuse,
transient -- but longer lived -- cores of the same mass, which will
not collapse to form stars ({\AA} Nordlund, private
communication). Thus although very low-mass prestellar cores might be
sufficiently rare and compact to have escaped detection, some of this
extensive population of non-prestellar cores should already have been
observed. However, to date there are very few candidate brown dwarf-mass cores
(Greaves 2005).

If, instead, most prestellar cores collapse and fragment in a
self-similar manner to form small-$N$ sub-clusters, the efficiency for
converting cores into stars needs to be somewhat higher,
$\sim\!50\%$. The observed binary statistics of Sun-like stars then
imply that a sub-cluster should only comprise 3 or 4 stars (in order
to match the overall binary frequency) and should have quite a wide
range of masses, $\sigma_{_{\ell og_{_{\footnotesize 10}}[M]}}\sim
0.6$ (in order to reproduce the observed low mass ratios; Goodwin \&
Kroupa 2005; Hubber \& Whitworth 2005). However, there is then no
explanation for the observed decline in binary frequency with
decreasing primary mass, or for any of the other systematic trends
with decreasing primary mass ({\it viz.} decreasing mean semi-major
axis, decreasing logarithmic spread of semi-major axes, increasing
mass ratio; e.g. Burgasser et al. 2006).

\section{Binary disruption}\label{SEC:BIDI}

In this section we propose a new mechanism for the formation of 
brown dwarfs. We hypothesise that brown dwarfs form as distant companions to
low-mass stars, in particular M-dwarfs. Such systems are then readily
disrupted by the mild perturbations of passing stars at relatively large distances 
(a few hundred AU or more). The result is a population of single brown 
dwarfs and low-mass hydrogen-burning stars. 

The disruption of these systems will usually be due to a small 
velocity perturbation induced by a passing star within the overall cluster 
(rather than the close and violent interactions in a small-$N$ system 
invoked by the standard ejection hypothesis). Also, as the disruption is gentle,
both the brown dwarfs and the low-mass hydrogen-burning stars are able
to retain their circumstellar discs. The relative velocities of the
components after disruption are low, and therefore there is little
spatial or kinematical difference between them and the more massive
stars in the same cluster. 

This picture avoids the problem of having
to create many brown dwarf-mass cores, as in the turbulence mechanism. It also
avoids the problems of giving some brown dwarfs a noticeably larger
velocity dispersion than the other stars in the cluster, stripping
circum-brown dwarf discs, and of hardening the surviving binary systems, as 
in the ejection mechanism.

\subsection{The formation of wide, low-mass binary systems}

This new mechanism for brown dwarf formation requires firstly that
brown dwarfs form as distant companions to low-mass hydrogen-burning
stars, and secondly that the resulting soft binary systems are gently
dissolved by tidal interactions with passing stars.

The simplest way to form such soft binaries is by disc
fragmentation. Rafikov (2005), Matzner \& Levin (2005) and Whitworth
\& Stamatellos (2006) have shown that it is very difficult for
low-mass companions to form by gravitational instability in the inner
parts of circumstellar discs. This is because the inner parts of a
disc are too warm, and too strongly irradiated by the central star,
for proto-fragments to cool radiatively on a dynamical
timescale. Thus, even if the disc is Toomre unstable (Toomre 1964),
proto-fragments which try to condense out undergo an adiabatic bounce
and are then sheared apart (Gammie 2001). However, low-mass companions
can condense out at larger radii,
\begin{eqnarray}
R&\;\ga\;&150\,{\rm AU}\,
\left(\frac{M_{_\star}}{{\rm M}_{_\odot}}\right)^{1/3}\,,
\end{eqnarray}
because here optical depths are lower, stellar irradiation is weaker,
and proto-fragments are therefore able to cool radiatively fast enough
to condense out (Whitworth \& Stamatellos 2006). The minimum mass for
condensations in the outer disc is
\begin{eqnarray}
M_{_{\rm MIN\,D}}&\;\sim\;&0.003\,{\rm M}_{_\odot}\,
\left(\frac{M_{_\star}}{{\rm M}_{_\odot}}\right)^{-1/4}\,\left(\frac{L_{_\star}}{{\rm L}_{_\odot}}\right)^{3/8}\,,
\end{eqnarray}
so such condensations should usually end up with brown dwarf masses 
($<0.075\,{\rm M}_{_\odot}$).

We conclude that soft, low-mass binary systems should form routinely,
provided that low-mass stars acquire sufficiently massive and extended
circumstellar discs to spawn brown dwarfs in the manner described above
(see Whitworth \& Stamatellos (2006) for further details). From a
theoretical perspective, this is at least plausible: observational
estimates of the mean specific angular momentum, $h$, in a core
(Bodenheimer 1995; his Fig. 1) give values $h\ga 10^{21}\,{\rm
cm}^2\,{\rm s}^{-1}$, and, if deposited in orbit around a star with
mass $M_{_\star} \sim 0.3\,{\rm M}_{_\odot}$, this material should end
up at radius
\begin{eqnarray}
R&\;\sim\;&\frac{h^2}{G\,M_{_\star}}\;\,\ga\,\;100\,{\rm AU}\,.
\end{eqnarray}
The fact that few such discs are observed can be attributed to the
likelihood that they fragment on a dynamical timescale. At $R\sim
100\,{\rm AU}$, this dynamical timescale is $t_{_{\rm ORBIT}}\sim
2000\,{\rm years}$, and therefore such discs should be very rare. The
resulting binary systems are expected to last somewhat longer ($\sim
5\,{\rm Myr}$ in a typical cluster; see Section 4.2), and indeed there
are several known binary systems in which the primary is a K- or
M-type star, and the secondary is a brown dwarf (Burgasser et al. 2005).

\subsection{The disruption of wide, low-mass binary systems}

In the mass range $M_{_\star}>0.5\,{\rm M}_{_\odot}$, not only is the
multiplicity significantly higher for pre-main sequence objects in
young clusters than for mature stars in the field (see Goodwin \&
Kroupa 2005; Goodwin et al. 2007; Duch\^ene et al. 2007; and
references therein), but also most of the excess is at large
separations $\ga 100\,{\rm AU}$ (compare Mathieu 1994, Patience et
al. 2002 and Duch\^ene et al. 2007 with Duquennoy \& Mayor 1991). As
first pointed out by Kroupa (1995a,b), this difference is most easily
explained by the dynamical destruction of binaries in their birth
clusters. Wide and/or low-mass binaries are especially susceptible to
destruction. Hence, starting with an initial binary fraction of unity,
Kroupa was able to recover the lower field binary fraction, the
separation distribution, and the dependence of these parameters on
primary mass, by simulating dynamical interactions in a representative
cluster. There would seem to be no other way to reconcile the high
binary fraction for pre-main sequence stars, as compared with the
field (Goodwin et al. 2007).

For lower masses, $M_{_\star}<0.5\,{\rm M}_{_\odot}$, the binary
fraction in young clusters is poorly constrained; but in the field,
the binary fraction for M-dwarfs appears to be significantly lower
($\sim 30\pm10\%$; Fischer \& Marcy 1992) than for G-dwarfs ($\sim
50\pm10\%$; Duquennoy \& Mayor 1991). This has lead Lada (2006) to
suggest that, whereas more massive stars usually form in multiples,
less massive ones usually form as singles. However, since it is the
low-mass systems which are most readily disrupted, and since
disruption must already be invoked for higher-mass stars to explain
the systematic shift in binary fraction between clusters and the
field, it seems much more likely that low-mass stars are also born in
multiple systems and are then selectively disrupted by dynamical
interactions.

To estimate the timescale for disruption, in a cluster, consider a
binary system with components of mass $M_{_1}$ and $M_{_2}$,
semi-major axis $a$, and hence binding energy $E_{_{\rm
BINDING}}=GM_{_1}M_{_2}/2a$. Using the impulse approximation (Binney
\& Tremaine 1987), a star of mass $M_{_\star}$ passing at distance $D$
with velocity $v$ delivers a tidal impulse
\begin{eqnarray}
\Delta E_{_{\rm TIDAL}}&\;\sim\;
&\frac{4\,G^2\,M_{_\star}^3\,a}{3\,v^2\,D^3}\,.
\end{eqnarray}
The system will be unbound if $\Delta 
E_{_{\rm TIDAL}}>E_{_{\rm BINDING}}$, i.e. if
\begin{eqnarray}
D&\;<\;&\left(\frac{G\,a^2}{3\,v^2\,M_{_1}\,M_{_2}}\right)^{1/3}\,
2\,M_{_\star}\,.
\end{eqnarray}
Substituting $a\sim 100\,{\rm AU}$, $v\sim 2\,{\rm km}\,{\rm s}^{-1}$,
$M_{_1}\sim 0.3\,{\rm M}_{_\odot}$, $M_{_2}\sim 0.03\,{\rm
M}_{_\odot}$ and $M_{_\star}\sim {\rm M}_{_\odot}$, this gives
$D<2000\,{\rm AU}$. In a cluster with density $n_{_\star}\sim
2000\,{\rm pc}^{-3}$, the rate for such impulsive interactions is
$\sim 0.2\,{\rm Myr}^{-1}$. In other words, a binary with $M_{_1}\sim
0.3\,{\rm M}_{_\odot}$, $M_{_2}\sim 0.03\,{\rm M}_{_\odot}$ and $a\sim
100\,{\rm AU}$ has a life expectancy of $\sim 5\,{\rm Myr}$ in such a
cluster. (Aguilar \& White (1985) have shown that, even in situations
like this, where the orbital velocities of the binary components are
comparable with the velocity dispersion of the stars in the cluster,
the impulse approximation gives reliable results.)

We conclude that tidal disruption of wide low-mass binaries will
result in a growing population of single brown dwarfs and low-mass
hydrogen-burning stars. The gentle nature of the disruption will allow
the separated binary components to retain circumstellar discs, and
hence to sustain accretion and outflows. The disruption velocities
will be comparable with the orbital velocities ($\la 1\,{\rm km}\,{\rm
s}^{-1}$), and hence less than -- or on the order of -- the velocity
dispersion in the cluster. Consequently, brown dwarfs will have
essentially the same kinematics and the same spatial distribution as
more massive stars.

Bouy et al. (2006) have tentatively identified a population of wide,
low-mass binaries in the Upper Scorpius OB Association, with
separations between $100\,{\rm AU}$ and $150\,{\rm AU}$, as required
by our model. They note that no such population is observed in the
Pleiades, and suggest that binary properties may change with
environment. An alternative explanation is that in the Pleiades, with
an age $\sim 100\,{\rm Myr}$, the primordial population of wide,
low-mass binaries has all been disrupted; whereas in Upper Sco OB,
with an age $\sim 5\,{\rm Myr}$ and relatively low density, a 
significant fraction of the primordial population survives.  

\subsection{The binary properties of brown dwarfs and very low-mass 
hydrogen-burning stars}

In the field, the binary frequency is estimated to be $20\pm 10\%$ for
brown dwarfs (L dwarfs), as compared with $30\pm10\%$ for very
low-mass hydrogen-burning stars (M dwarfs) and $50\pm10\%$ for
Sun-like stars (G dwarfs) (e.g. Burgasser et al. 2006; Basri \&
Reiners 2006; Duquennoy \& Mayor 1991; Fischer \& Marcy 1992; Luhman
et al. 2007; Reid et al. 2006). There is some evidence to suggest that
the binary frequency for brown dwarfs may be significantly higher in
clusters, say $40\pm10\%$ (Pinfield et al. 2003; Maxted \& Jeffries
2005; Chappelle et al. 2006; Montagnier et al. 2006), but this claim
awaits confirmation. As primary mass increases, the mean binary
separation also increases, and the mean mass ratio, $q\equiv
M_{_2}/M_{_1}$, decreases (e.g. Lucas et al. 2005; Kraus et al. 2006).

As an indication of how our model might relate to these statistics,
suppose that the primordial binary fraction for M dwarfs is $100\%$,
so a cluster might start with 80 primordial M-dwarf binaries. We then
require that 30 of these primordial binaries survive (predominantly
with M-dwarf secondaries), 20 others with M-dwarf secondaries are
disrupted, and 30 with brown-dwarfs secondaries are also
disrupted. This leaves 30 M-dwarf binaries, 70 M-dwarf singles, and 30
brown-dwarf singles -- hence an M-dwarf binary fraction of
$\sim\!30\%$ and a ratio of 10 M dwarfs to every 3 brown dwarfs. In
total, 50 primordial binaries (i.e. $\sim\!60\%$ of the original 80)
have been disrupted, and these will tend to be the wider ones. Whilst
the above reckoning is not intended to be definitive, we note that it
is consistent with the over-abundance of wide binaries ($\ga 100\,{\rm
AU}$) observed in young populations, relative to the field (Patience
et al. 2002; in particular, their Fig. 4). It is also consistent with
the distribution of mass ratios observed in field M-dwarf binaries by
Fischer \& Marcy (1992), which is noticeably flatter than for G-dwarfs
(provided one excludes very close systems; Mazeh et al. 1992) and
therefore favours less extreme mass ratios in the evolved field
population.

\subsection{Close brown dwarf-brown dwarf binaries}

There appears to be a significant population of close ($<20\,{\rm
AU}$) brown dwarf-brown dwarf binaries (Basri \& Mart\'in 1999; Close et al. 2003;
Pinfield et al. 2003; Maxted \& Jeffries 2005; Burgasser et al. 2006;
Luhman et al. 2007). It is unlikely that such binaries are formed by
ejection, since brown dwarfs and low-mass hydrogen-burning stars
formed by ejection are almost always single. Equally, a brown-dwarf
secondary is unlikely to form by fragmentation of a circumstellar disc
around a brown-dwarf primary. Firstly, if the disc is sufficiently
massive to fragment, then accretion onto the primary is likely to push
its mass above the hydrogen-burning limit. Secondly, fragmentation is
precluded {\it in the inner parts} ($< 100\,{\rm AU}$) of a
circumstellar disc, because the gas is too warm and too strongly
irradiated by the central star (Whitworth \& Stamatellos 2006), and
therefore disc fragmentation cannot produce close binaries. Close
brown dwarf-brown dwarf binaries may form by secondary fragmentation, due to the
dissociation of molecular hydrogen (see Goodwin et al. 2007, Whitworth
et al. 2007). For example, Whitworth \& Stamatellos (2006) suggest
that low-mass fragments condensing out {\it in the outer parts} ($\ga
100\,{\rm AU}$) of a circumstellar disc -- as we are suggesting here
-- may subsequently undergo secondary fragmentation to produce close
brown dwarf-brown dwarf binaries. Gentle disruption can then sometimes 
separate the close brown dwarf-brown dwarf binary from the star at the 
centre of the disc without destroying the close brown dwarf-brown
dwarf binary itself, thereby populating the field with close brown 
dwarf-brown dwarf binaries. This scenario would explain the observation
that brown dwarfs which remain in wide orbits around more massive
stars are -- {\it modulo} small number statistics -- much more likely
to be in a close binary system with another brown dwarf than are brown
dwarfs in the field (Burgasser et al. 2005).

\subsection{Possible observational tests of this model}

This scenario requires that brown dwarfs form in massive, but
  short-lived discs around low-mass stars.  Such a disc is expected
  to fragment on a dynamical timescale, and so its lifetime is much
  shorter than that of a class 0 object (see e.g. Ward-Thompson et
  al. 2007).  Observations of such discs will therefore be hard. 
For example, the dynamical timescale for a disc with diameter 
$\sim 100\,{\rm AU}$ around a primary with mass 
$M_{_1}\sim 0.3\,{\rm M}_{_\odot}$ is $\sim 3000\,{\rm years}$. If we 
assume that the brown dwarf formation rate is $\sim 0.2/{\rm year}$, 
and that brown dwarf formation is distributed uniformly over the Galactic 
disc out to $\sim 10\,{\rm kpc}$, then there is on average only one such 
disc within $\sim 400\,{\rm pc}$, and it is also presumably a deeply 
embedded, low-luminosity source. ALMA should have no difficulty 
resolving such a source, but finding it and collecting sufficient 
photons to identify it unambiguously in a crowded and confused 
region may be more problematic.

The timescale and efficiency of the disruption of wide, low-mass 
binaries should depend on the density of the parent cluster. 
Therefore we would expect loose associations to still contain 
undisrupted systems.  Such a population is indeed observed in the 
upper Sco OB association (Bouy et al. 2006), but not in Taurus (Kraus 
et al. 2006). However, we note that most star formation 
in Taurus appears to occur in fairly dense subgroups whose density 
might be sufficient to disrupt a low-mass, wide binary
population.  In addition, we note that Taurus appears 
to be deficient in M-dwarfs {\em  and} brown dwarfs 
relative to a `standard' IMF (see Goodwin et al. 2004c 
and references therein), and this lack of M-dwarfs may explain the
lack of brown dwarfs relative to `typical' regions.  We plan to 
conduct a series of
$n$-body simulations to test the efficiency of disruption in different
environments.

\section{Conclusions}\label{SEC:CONC}

We propose that there exists a primordial population of wide low-mass
binaries, most of which are rapidly but gently disrupted by tidal
encounters in clusters, to produce the observed population of single
brown dwarfs and low-mass hydrogen-burning stars. This primordial
population corresponds to the excess of wide binaries found amongst
pre-main sequence stars (e.g. Mathieu 1994, Patience 2002). Such wide,
low-mass binaries are highly susceptible to disruption by passing
stars, and are unlikely to survive long in the dense environment of a
young cluster. However, the disruption of such binaries will normally
be very gentle, allowing the individual components (brown dwarfs and
low-mass hydrogen-burning stars) to retain circumstellar discs, and
hence to sustain accretion and outflows. For the same reason, the
peculiar velocities of the components after disruption will be low,
and therefore they will be hard to distinguish from the more massive
stars in the cluster on the basis of their velocity dispersion or
spatial distribution.

This model produces a predominantly single population of brown dwarfs
and low-mass hydrogen-burning stars, with a continuum of statistical
properties, both across the hydrogen-burning limit, and towards higher
masses. It obviates both the need to produce very low-mass prestellar
cores, as in the pure turbulence theory, and the large peculiar
velocities generated by the ejection mechanism.

Lada (2006) has suggested, based on the low {\em field} binary fraction of
M-dwarfs, that most stars (ie. M-dwarfs) form as single objects. The 
model we have presented in this paper, however, implies that 
{\em all} stars and brown dwarfs form within binary and multiple 
systems.  In this case, any theory of star formation must produce
multiple systems as the norm.

\begin{acknowledgements}
The authors would like to thank Adam Kraus for useful discussions.  
SPG was supported for part of this work by a UKAFF Fellowship. APW's 
work is supported by PPARC (Ref. PPA/G/O/2002/00497).  
\end{acknowledgements}

\end{document}